\documentstyle[prc,aps,preprint]{revtex} 
\tightenlines
\begin{document}
\draft
\def\pmb#1{\setbox0=\hbox{#1}%
     \kern-.025em\copy0\kern-\wd0
      \kern.05em\copy0\kern-\wd0
       \kern-.025em\raise.0433em\box0}
\def\btau{\pmb{$\tau$}}
\def\bsigma{\pmb{$\sigma$}}
\def\bdiamond{\pmb{$\diamond$}}
\def\bbdiamond{\pmb\bdiamond}
\def\hat{\widehat}
\def\omeg{\bmatdieci\char '41}
\date{\today}
\title
{Fock exchange terms in non--linear Quantum Hadrodynamics}
\author{
V.Greco${}^{{1}}$, F.Matera${}^{{2}}$, M.Colonna${}^{{1}}$, 
M.Di Toro${}^{{1}}$ and G.Fabbri${}^{{2}}$}

\address
{$^{1}$Laboratorio Nazionale del Sud, Via S. Sofia 44,
I-95123 Catania, Italy\\
and Universit\`a degli Studi di Catania\\}

\address
{$^{2}$Dipartimento di Fisica, Universit\`a degli Studi di Firenze\\
and Istituto Nazionale di Fisica Nucleare, Sezione di
Firenze,\\
L.go E. Fermi 2, I-50125, Firenze, Italy}
\maketitle

\begin{abstract}
We propose a method to introduce Fock term contributions in relativistic 
models of fermions coupled to mesons, including 
self-interactions for the mesonic fields. We show that effects
on equilibrium properties and on the dynamical response of the fermionic system
can be consistently accounted for. 
Some implications on equilibrium properties of asymmetric nuclear matter
are discussed.
In particular an indication is emerging for a reduced contribution of
charged mesons to the symmetry term of the nuclear equation of state
around normal density.
\end{abstract}


\vskip 1.0cm

\section{ Introduction}

Since the pioneering work of Walecka \cite{Wa74}, the relativistic 
field model of hadrons, called Quantum Hadrodynamics (~QHD~),  
with its developments and extensions, has been widely used in 
investigations of nuclear systems, see the extensive reviews  
\cite{Se86,Se92,Se97,Ri96}. 
This model can be regarded as an {\it effective field theory} 
\cite{Se97,Br96,Fr99} at the nuclear scale, and it is 
based on a relativistic effective Lagrangian which includes strongly 
interacting nucleon and meson fields. The success of QHD has been due 
to the fact that it provides a realistic description of bulk 
properties  both of infinite nuclear matter and of finite nuclei 
\cite{Se86,Se92,Se97,Ri96}, even with a minimal number of 
meson fields, the 
neutral 
$\sigma$ and $\omega$ fields and the charged $\rho$ field.

The usual procedure to deal with theories which contain baryons 
interacting with mesons consists of eliminating the meson fields 
in favour of the fermionic degrees of freedom. This can be done 
by exploiting the equations of motion for the meson fields. In this way
the physical system can be ultimately described in terms of nucleons 
interacting through the exchange of virtual mesons. 

It is important to notice that, once
the Fock exchange terms of the nucleon--nucleon interaction are taken into 
account, the Walecka model can reproduce 
some basic properties of 
nuclear matter by means of the 
exchange of the scalar $\sigma$ and vector $\omega$ fields alone 
(~ see Refs. quoted in \cite{Se86,Se92,Se97}~). 
Including the Fock terms also the whole variety of processes in nuclear 
dynamics arising from the fermionic intrinsic degrees of freedom 
(~spin and isospin~), can be  accounted for\cite{De91,Ma94,Be93}. 
In principle, if the model is constructed self--consistently, 
the introduction of further meson fields with the corresponding 
constants for the coupling to nucleons appears to be less
necessary. We will elaborate in detail on this point.

However, in its original version, the QHD model suffered 
from a severe deficiency, namely it was not able to give a reasonable 
value for the nuclear compressibility modulus
and the effective mass.
This unpleasant 
drawback has been cured by introducing  self--interactions 
of the scalar $\sigma$ field up to fourth order \cite{Bo77,Bo89}. 
Moreover, within both the modern scheme of effective field 
theories \cite{Fu96,Mu96} and hadronic chiral models 
\cite{Fu97} 
a non--linear self--interaction for the scalar meson 
emerges in a natural way. 

Calculations based on non--linear models of QHD have been so far 
performed in the relativistic mean field approximation (~RMF~), which 
essentially corresponds to the Hartree approximation. With the 
inclusion of these higher order terms in the Lagrangian, 
the evaluation of the exchange terms of the nucleon--nucleon interaction 
represents a very difficult task. This is due to the non--linearity 
of the resulting equation of motion for the meson field. 
Although the RMF approximation to QHD with non--linear terms has achieved 
a large success in reproducing quantitatively many observables, in 
particular for finite nuclei \cite{Ri96}, the inclusion 
of the Fock terms would be highly desirable. In fact, these terms 
originate from the correlations due to the Fermi--Dirac statistics, therefore 
they are related to a genuine quantum effect, which in general cannot 
be neglected in studying a many--body system.

In this paper, we present a procedure which allows to evaluate the 
contribution of the Fock terms in a truncation scheme. The result 
is an expansion of the Fock terms in a $1/N$ series, 
where $N$ is the multiplicity of the fermionic intrinsic degrees 
of freedom (~$N=4$ for nucleons~). 
In this sense the approach suggested here is not of perturbative
nature in the coupling constants, which are always accounted for at
all orders.

We illustrate our method in 
the case of infinite nuclear matter. We show that both equilibrium 
properties and dynamics can be self--consistently treated at the 
same degree of approximation. 

We would like to remark that our treatment of the Fock terms 
is not restricted to QHD models. It can be applied in general, 
most likely with non trivial consequences, to models involving 
fermions coupled to bosons. Among these, 
we mention the Quark Meson Coupling model \cite{Gu88}, 
where nuclear matter is described as a collection of composite 
nucleons with meson fields directly coupled to quarks 
\cite{Sa94,Bl96,Li96}.
Outside a strictly nuclear context, we 
mention also the Linear Quark Meson model, which represents an 
{\it effective} description of QCD below the mesonic compositeness 
scale (~$\sim 600 MeV$~) (~see the reviews \cite{Ju97} and \cite{Be99}~). 
This model provides a hybrid description of QCD in terms of quarks and 
mesons. 

\section{Treatment of Fock Terms in a Kinetic Approach}

The aim of the present paper is to illustrate a procedure which allows to take
into account the self--interactions of meson fields within a Hartree--Fock 
scheme. For simplicity, and also to focus on the new procedure, we limit
ourselves to consider only the minimal number of mesons which is 
sufficient to reproduce some basic properties of nuclear matter. Therefore
our starting point is the QHD-I $\sigma-\omega$ model \cite{Se86}, including 
self--interaction terms for the scalar $\sigma$ field \cite{Bo77,Bo89}. 
The Lagrangian density for this model is given by:
\begin{eqnarray}
{\cal L} = {\bar {\psi}}[\gamma_\mu(i{\partial^\mu}-{g_V}{\cal V}^\mu)-
(M-{g_S}\phi)]\psi + {1 \over 2}({\partial_\mu}\phi{\partial^\mu}\phi
- {m_S}^2 \phi^2) \nonumber \\
- {a \over 3} \phi^3 - {b \over 4} \phi^4 - {1 \over 4} W_{\mu\nu}
W^{\mu\nu} + {1 \over 2} {m_V}^2 {\cal V}_\mu {{\cal V}^\mu}\, ,
\end{eqnarray}
where $W^{\mu\nu}(x)={\partial^\mu}{{\cal V}^\nu}(x)-
{\partial^\nu}{{\cal V}^\mu}(x)~.$ Here $\psi(x)$ 
denotes the 8(spin and isospin)-component 
nucleon field, the scalar ($\phi(x)$) and the vector (${{\cal V}^\nu}(x)$) 
boson fields are associated with the scalar 
($\sigma$) and vector (~$\omega$~) mesons, respectively ($\hbar$ = $c$ = 1).  
 
Fock contributions naturally appear in the source terms of the 
equation of motion of the fields. Here we will study the related effects 
on the corresponding kinetic equations, in particular for the
fermion field. This will allows us to derive the contributions
of exchange terms on static and dynamical properties of the
interacting nuclear system in a self-consistent way, at the
same level of approximation.

\subsection{The Wigner transform formalism}

In agreement with the Relativistic Mean Field ($RMF$) picture
of the $QHD$ model
we focus our analysis on a description of 
the many-body nuclear system in terms of one--body dynamics. 
Therefore, it is convenient to use the Wigner transform 
of the one--body density matrix of the fermionic field. 
The one--particle Wigner function is defined as:
$$[{\widehat F}(x,p)]_{\alpha\beta}=
{1\over(2\pi)^4}\int d^4Re^{-ip\cdot R}
<:\bar{\psi}_\beta(x+{R\over2})\psi_\alpha(x-{R\over2}):>~,$$
where $\alpha$ and $\beta$ are double indices for spin and isospin. 
The brackets denote statistical averaging and the colons denote 
normal ordering. 
According to the Clifford algebra, 
the Wigner function ${\hat F}(x,p)$ can be decomposed as:
$${\hat F}(x,p)= F(x,p)+{\gamma_\mu}{F^\mu}(x,p)+{1\over2}
{\sigma_{\mu\nu}}F^{\mu\nu}(x,p)
$$
\begin{equation}
+\gamma^5A(x,p)-{\gamma^5\gamma_\mu}{A^\mu}(x,p)~.
\end{equation}
The integral over $p$ of the Wigner function 
$$[{\widehat F}(x)]_{\alpha\beta}=\,
\int d^4p\,[{\widehat F}(x,p)]_{\alpha\beta}=\,
<:\bar{\psi}_\beta(x)\psi_\alpha(x):>$$
is related to the various densities with their specific transformation 
properties both in ordinary space and in isospin space (~isoscalar 
and isovector~).
For instance the scalar and current isoscalar 
densities are given by: 
$$\rho_S(x)=\,<:{\bar \psi(x)}\psi(x):>=\,Tr{\widehat F}(x) = 8~F(x)\, ,$$
$$j^{\mu}(x)=\,<:{\bar \psi(x)}\gamma^{\mu}\psi(x):>=
\,Tr\gamma^{\mu}{\widehat F}(x) = 8~F^{\mu}(x)\, .$$

The equation of motion for the Wigner function can be derived from 
the Dirac equation by using standard procedures 
(see e.g. Refs.\cite{degr80,hak78}):
\begin{eqnarray}
{i\over2}{\partial_\mu}[\gamma^\mu{\hat F}(x,p)]_
{\alpha\beta}+
p_\mu[{\gamma^\mu}{\hat F}(x,p)]_{\alpha\beta}-
M{\hat F}_{\alpha\beta}(x,p) \nonumber \\
-g_V{1\over(2\pi)^4}\int d^4Re^{-ip\cdot R}
<:\bar{\psi}_\beta(x_+){\gamma^\mu_{\alpha\delta}}\psi_\delta(x_-)
{\cal V}_\mu(x_-):> \nonumber \\
+g_S{1\over(2\pi)^4}\int d^4Re^{-ip\cdot R}
<:\bar{\psi}_\beta(x_+)\psi_\alpha(x_-)\phi(x_-):>=0~,
\end{eqnarray}
with $x_+=x+{R\over2}$ and $x_-=x-{R\over2}$.

\subsection{The truncation scheme for exchange contributions}

The statistical averages of the fermionic and mesonic operators 
in the above equation will now be treated within a Hartree--Fock 
(~HF~) approximation. 
In order to take into account the contribution 
of exchange terms in a manageable way we assume, as a 
basic approximation, that in the equations of motion for the meson 
fields the terms containing 
derivatives can be neglected with 
respect to the mass terms. Then, the meson   
field operators are connected to the operators of the nucleon 
scalar and current densities by the following equations:
\begin{eqnarray}
{\widehat{\Phi}/f_S} + A{\widehat{\Phi}^2}
+ B{\widehat{\Phi}^3}&=\bar\psi(x)\psi(x)={\widehat\rho}_S(x)
~,\nonumber \\
{\widehat V}^\mu(x)&=f_V\bar\psi(x){\gamma^\mu}\psi(x)={\widehat 
j}_\mu(x)~,
\end{eqnarray}
where ${\widehat \Phi} = g_S \phi$, 
${\widehat V^{\mu}} = g_V {\cal V}^{\mu}$ and
$f_S = (g_S/m_S)^2$, $f_V = (g_V/m_V)^2$,
$A = a/g_S^3$, $B = b/g_S^4$.  
\par
This approximation is not justified for light mesons, like pions.
The inclusion of self--interaction terms of the pionic field is a very
difficult task and a different approximation scheme is needed. However
in this case a perturbative expansion
in the pion-nucleon coupling constant seems to be reasonable \cite{Hor83}.
Moreover it has been shown that
the inclusion of pions does not change qualitatively the description of
nuclear matter around normal conditions \cite{Hor83}.
  
The next step will be the elimination of the meson 
fields in Eq.(3) by means of Eq.(4). As far as the term containing 
the vector field is concerned, by a simple substitution one 
can obtain a statistical average of only fermionic 
field operators \cite{De91}. 
Instead, because of the non--linear coupling of the $\sigma$ 
field to the fermionic scalar density, the elimination of 
the scalar field in the last term of Eq.(3) requires a 
particular procedure.   

We consider the expansion 
of the field operator ${\widehat \Phi}$ in terms of the 
scalar density operator ${\widehat {\rho}_S} $: 
${\widehat \Phi} = \sum_n \chi_n {\widehat {\rho}_S}^n$. Then, in 
the HF approximation , by means of the Wick's theorem 
the average in the last term of Eq.(3) can be expressed 
as: 
\begin{eqnarray}
<:{\bar \psi}_\beta(x_+)\psi_\alpha(x_-){\widehat \Phi}(x_-):> =
<:{\bar \psi}_\beta(x_+)\psi_\alpha(x_-):><:{\widehat \Phi}(x_-):> 
+ <:{\bar \psi}_\beta(x_+)\psi_\gamma(x_-):>
\nonumber \\
\times 
\sum_{k=1}{1\over {k!}} (-1)^k
[{\widehat F}^k(x_-)]_{\alpha\gamma}\,<:{{d^k{\widehat \Phi}(x_-)} 
\over {d{\widehat \rho}^k_S(x_-)}}:>\, . 
\end{eqnarray}
The statistical averages of ${\widehat \Phi}(x)$ and its 
derivatives are given by series of $<:{\widehat {\rho}_S}^n(x) :>$, 
which in turn can be expanded as: 
\begin{equation}
<:{\widehat \rho_S}^n(x) :> = \rho_S^n(x) - \sum_{k=2}{1\over {k!}}
(-1)^k Tr {\widehat F}^{k}(x)
{{d^{k}\rho_S^n}(x) \over {d\rho_S^{k}(x)}}\, . 
\end{equation}
As a consequence, we obtain for $<:{\widehat\Phi}(x):>$ the expansion: 
\begin{equation}
<:{\widehat \Phi}(x) :> = \Phi(x) - \sum_{k=2}{1\over {k!}}
(-1)^k Tr {\widehat F}^{k}(x)
{{d^{k}\Phi}(x) \over {d\rho_S^{k}(x)}}\, , 
\end{equation}
where  $\Phi(x)=\,\sum_n \chi_n\rho_S^n(x)$, and similar expressions 
for $<:d^{k}{\widehat \Phi}(x)/d^{k}{\widehat \rho}_S(x):>$.
The expansions of Eqs.(5),(6) and (7) allow to settle a truncation
scheme. The parameter to fix the approximation order is given 
by the number of factors ${\widehat\rho}_S(x)=\bar\psi(x)\psi(x)$ 
which are broken in writing Eq.(5) with the expansion (7). 
For each decoupling there is a trace operation less, leading each time to 
a quenching factor of $1/8$. Moreover in our case the coefficients 
$A$ and $B$ in Eq.(4), to be determined by fitting the bulk properties 
of nuclear matter, should be small  \cite{Fu96} . 
Therefore, the derivatives $d^k\Phi(x)/d\rho_S^k(x)$ become smaller and 
smaller as $k$ increases. 

The expansion Eq.(5) actually implies that contributions of nucleons of 
the Dirac sea 
are neglected. This approximation will allow us to
derive a closed equation for the Wigner function containing only finite
quantities. We discuss this point at the end of the section.

We consider exchange corrections only up to 
the next--to--leading term, with respect to the Hartree contribution. 
In this approximation  
Eq.(5) assumes the form: 
\begin{eqnarray}
<:{\bar \psi_\beta}(x_+)\psi_\alpha(x_-){\widehat \Phi}(x_-):> = 
<:{\bar \psi_\beta}(x_+)\psi_\alpha(x_-):> (\Phi(x_-) - {1\over 2}
Tr {\widehat F}^{2}(x_-)
{{d^{2}\Phi}(x_-) \over {d\rho_S^{2}(x_-)}}) 
\nonumber \\
- <:{\bar \psi_\beta}(x_+)\psi_\gamma(x_-):>
[{\widehat F}(x_-)]_{\alpha\gamma}
{{d\Phi}(x_-)
\over {d\rho_S}(x_-)},
\end{eqnarray}
where the function $\Phi(x)$ obeys the equation: 
${\Phi(x)}/f_S + A{\Phi^2}(x)
+ B{\Phi^3}(x) =\rho_S(x).$ 

According to the procedures usually followed in the derivation of kinetic
equations (Refs.\cite{De91,degr80}) (semi-classical approximation), 
we assume that the densities 
$<:{\bar \psi_\gamma}(x_-)\psi_\alpha(x_-):>$ and the field 
$\Phi(x_-)$ are slowly varying functions, and we retain only 
the first order term in their Taylor expansion at the point $x$. 
This is essentially the same kind of assumption 
previously done for the meson fields and expressed by Eq.(4).
In fact because of the small Compton wave--lengths 
of the heavy mesons $\sigma$ and $\omega$, 
retardation and finite range  effects can be neglected. 

Within the approximations outlined above Eq.(3) becomes 
$$
{i\over2}{\partial_\mu}{\gamma^\mu}{\hat F}(x,p)+
\gamma^\mu (p_\mu-f_Vj_\mu(x)){\hat F}(x,p)-
\biggl(M-\Phi(x)+
{1\over 2}Tr {\widehat F}^{2}(x)
{{d^{2}\Phi(x)} \over {d\rho_S^{2}(x)}}
\biggr){\hat F}(x,p) 
$$
$$
+ {i\over2}\Delta
\{f_Vj_\mu(x)\gamma^\mu-\Phi(x) +
{1\over 2}Tr {\widehat F}^{2}(x)
{{d^{2}\Phi(x)} \over {d\rho_S^{2}(x)}} 
\}{\hat F}(x,p)
$$
\begin{equation}
-f_V[({i\over2}\Delta-1)\gamma^\mu{\hat F}(x)\gamma_\mu{\hat F}(x,p)]
+[({i\over2}\Delta-1)({{d\Phi(x)} \over {d\rho_S(x)}}
{\hat F}(x)){\hat F}(x,p)]=0~,
\end{equation}
where $\Delta={\partial_x}\cdot{\partial_p}$, with $\partial_x$ acting only 
on the first term of the products. 

This is the kinetic equation for the matrix ${\hat F}(x,p)$ in mean--field 
dynamics. In addition to the scalar and vector 
isoscalar fields, the exchange contributions, which are the terms 
containing the matrix ${\hat F}(x)$ in Eq.(9), may give rise to 
scalar, vector, tensor, 
pseudoscalar and pseudovector fields (~both isoscalar and isovector~), 
leading to completely new effects.\par
In Eq.(9) we have a general expression for the effective mass, 
including an isospin contribution. Since the
derivatives of mesonic fields have been neglected, the effective mass 
does not present a
momentum dependence. \par
A quantity of interest in the study of nuclear dynamics is the statistical
average of the canonical energy-momentum density tensor. 
Since terms containing the derivatives of the meson fields are neglected
in our approximation, the tensor takes the form:
\begin{equation}
T_{\mu\nu}(x) = {i\over 2}{\bar \psi}(x)\gamma_\mu \partial_{\nu} \psi(x) 
+ [ U({\widehat \Phi}) 
 - {1\over {2f_V}}{\widehat V}_\lambda(x) {\widehat V}^\lambda(x)
]g_{\mu\nu},
\end{equation}
where $g_{\mu\nu}$ is the diagonal metric tensor and
$U({\widehat \Phi}) =  
{1\over 2}{\widehat \Phi}^2/f_S + A/3~ {\widehat \Phi}^3 + B/4~
{\widehat \Phi}^4$. 
According to the approximation introduced above, 
the energy--momentum density tensor, in the HF approximation, reads:
\begin{eqnarray}
<: T_{\mu\nu}(x) :> =  8\int d^4 p~ p_{\nu}F_{\mu}(x,p) + 
\{U(\Phi) - {f_V\over 2}
j_\lambda(x) 
j^\lambda(x)\}g_{\mu\nu} \nonumber \\
- {1\over 2}  
[Tr {\widehat F}^{2}(x)
{{d^{2}U(\Phi)}\over {d\rho_S^{2}(x)}} - f_V Tr(\gamma_\lambda{\widehat F(x)}
\gamma^\lambda {\widehat F}(x))]g_{\mu\nu}
\end{eqnarray}
The quantities in square brackets are the exchange contributions.
It is essential to note that Fock terms contain traces of powers of
${\widehat F}(x)$
that naturally may bring contributions from all the fields having 
tensorial properties consistent with the symmetry of the system.

>From Eq.(11) we obtain the energy density for symmetric nuclear
matter that in analogy to the Hartree case can be rewritten 
in the following form:

\begin{equation}
\epsilon= {<:T_{00}:>} =  
{4\over (2\pi)^3}\int {d{\bf p}}  E^*_p
\theta (p_F-|{\bf p}|) + 
{U(\Phi)}+
{1\over2} 
{\tilde f_S} {\rho_S}^2 + {1\over2} {\tilde f_V} {\rho_B}^2 ,
\end{equation}

where $E^*_p = \sqrt{|{\bf p}|^2 + {M^*}^2}$,
$\rho_B=j_0$ is the baryon density and $p_F$ is the Fermi momentum.
$M^*$ is the nucleon effective mass: 
$$
M^* = M - {\Phi} -({1\over2} f_V - {1\over8}{{d\Phi}\over{d\rho_S}})\rho_S
+ {1\over 16} (\rho_S^2 + \rho_B^2)
{{d^2\Phi}\over{d\rho_S^2}}.
$$
The scalar density $\rho_S$ 
is given by 
$$
\rho_S\,={4 \over (2\pi)^3}M^*\,\int {d{\bf p}\over E^*_p}
\theta (p_F-|{\bf p}|)\,.
$$
The density dependent coupling constants ${\tilde f_S}$ and ${\tilde f_V}$
are defined as it follows: 
$$
{\tilde{f_S}}={1\over2} f_V - {1\over8}({{d\Phi}\over{d\rho_S}} +
 \rho_S {{d^2\Phi}\over{d\rho_S^2}});
$$
\begin{equation}
{\tilde f_V}={5\over4} f_V + {1\over8}({{d\Phi}\over{d\rho_S}} -
 \rho_S {{d^2\Phi}\over{d\rho_S^2}}).
\end{equation}

To give some indications about the convergency of the truncation in Eq.(8),
we plot in Fig.1 the density behaviour of the coupling functions 
${\tilde f}_S$ and ${\tilde f}_V$, obtained using for $f_S$, $f_V$, 
$A$ and $B$ the parameterization given in chapter III.   
The dashed lines
represent the results obtained neglecting the terms
containing the second derivative of $\Phi$ in Eq.(13), while the full
lines correspond to the complete calculations (Eq.(13)). 
Up to densities of the order of $4~\rho_0$, it is possible to observe that,
for the parameterization considered, 
the inclusion of the second derivatives changes slightly the results.
Therefore in such a density domain one expects a even smaller 
contribution from the next terms of the expansion of Eq.(6), since they contain
higher order derivatives.

\par
In order to establish a link with previous works on the introduction of Fock
terms in QHD theories, we will consider only the linear term 
in the equation for the scalar 
field, Eq.(4), where this kind of calculations have been performed 
so far [10,24-27],\cite{newcomment}.
 
Our results are clearly reducing to the ones
reported in Ref.[8]. The single--particle energies and 
the nucleon effective mass are given by: 
\begin{equation}
\epsilon_p\,=E^*_p + f_V\rho_B+[({1\over 4}f_V+{1\over 8}f_S)\rho_B] = 
E^*_p + {\tilde f_V}\rho_B
\end{equation}
and 
\begin{equation}
M^*\,=M-f_S\rho_S-[({1\over 2}f_V-{1\over 8}f_S)\rho_S]\ = 
M - (f_S + {\tilde f_S})\rho_S,
\end{equation}
respectively, where 
now ${\tilde f_S}$ and ${\tilde f_V}$ of Eq.(13) are simply given by:
${\tilde f_S} = -{1\over 8} f_S + {1\over 2} f_V$ ,
${\tilde f_V} = {1\over 8} f_S + {5\over 4} f_V$ .
The energy density, $\epsilon = <:T_{00}:>$,  
is expressed by: 
$$
\epsilon\,={4\over (2\pi)^3}\int {d{\bf p}}E^*_p\theta (p_F-|{\bf p}|)
~+{1\over 2}f_S\rho_S^2+{1\over 2}f_V\rho_B^2
$$
$$
+{1\over 2}[({1\over 2}f_V-{1\over 8}f_S)\rho_S^2+ 
({1\over 4}f_V+{1\over 8}f_S)\rho_B^2] = 
$$
\begin{equation}
{4\over (2\pi)^3}\int {d{\bf p}}E^*_p\theta (p_F-|{\bf p}|)
~+{1\over 2}(f_S + {\tilde f_S})\rho_S^2+{1\over 2}{\tilde f_V}\rho_B^2,
\end{equation}

with the exchange contributions in Eqs.(14),(15) and (16) given by the terms 
in square brackets. \par
We remark that 
the same expressions for $E_p$, $M^*$ and $\epsilon$ 
can be obtained from the results of the HF calculations 
of Ref.\cite{Hor83} neglecting the square of the four--momentum 
transfer between two interacting nucleons with respect to the 
squared meson masses. 
Hence for linear models of 
QHD our approach is equivalent to the usual HF approximation, once
retardation and finite--range 
effects are neglected. 

The two--loop approximation for nuclear matter \cite{Fu89} was 
introduced to study the loop expansion 
scheme for QHD. 
According to this 
approximation the contribution of exchange diagrams to the 
energy density is evaluated using nucleon propagators in the 
Hartree approximation (i.e. at one loop level). 
Neglecting vacuum, retardation and finite range effects, the results of
Ref.[25]
can be reproduced by using a similar approximation 
in our approach. This amounts to neglect the exchange 
terms in the equation for the equilibrium Wigner function, and to 
retain the exchange contributions in the equation for the 
energy density, Eq.(16). Therefore, within
this approximation, the 
exchange terms in the nucleon self--energy are disregarded. 
However it should be noticed that their contribution to the energy density, 
i.e. 
the second term in the square bracket of Eq. (16), $\Delta\epsilon=
({1\over 4}f_V+{1\over 8}f_s)\rho^2_B$, is comparable to 
the exchange contributions that are included in \cite{Fu89} and hence it should
not be neglected .  A similar situation 
occurs in the self--consistency equation for the nucleon 
effective mass. \par


Within the formalism used in 
the present paper, 
vacuum effects could be taken into account by omitting the normal ordering 
and subtracting a vacuum term in the definition of the Wigner 
function \cite{hak78}. Such a subtraction does not eliminate all 
the divergent vacuum terms in the equations for the Wigner 
function and the energy--momentum density tensor, thus a 
renormalization procedure in necessary \cite{hak78}. 
Since, as stressed before, our approach is equivalent to the HF 
approximation, we should expect 
to get results analogous to those obtained in \cite{Fu89,Biese84}. 
A fully 
self--consistent HF description of QHD including vacuum effects 
is very complicated \cite{Biese84} and has not been achieved yet. 
On the other hand, two--loop corrections 
to the Hartree approximation give unnaturally large vacuum 
contributions to the energy of nuclear matter 
\cite{Fu89}. 
This would 
suggest to follow other paths for 
representing the vacuum dynamics. 
For a review on these 
points see also Ref.\cite{Se97}. 

However, it is generally recognized that
the RMF approximation, where negative energy states of nucleons 
are not considered, 
is very successful in describing 
properties both of nuclear matter and of finite nuclei, once
self--interaction terms of the
$\sigma$--field  are included. 
The approach presented here can be considered as an 
extension of the RMF approximation to include exchange 
terms on the same basis.\par

\section{Influence of Fock Terms on the Nuclear Equation of State}

The study of the nuclear Equation of State ($EOS$) can be performed 
by evaluating the statistical average of the energy--momentum 
tensor, Eq. (11).\par 
Specializing to the case of symmetric nuclear matter, 
we have tested that our procedure to introduce the Fock terms
leads to a thermodynamically consistent theory. 
The equilibrium Wigner function together with the single--particle 
energy spectrum can be obtained following the scheme of Ref.\cite{De91}. 
We have found that  the relation between pressure and energy density: 
\begin{equation}
P =\,{\rho_B}{{d\epsilon}\over{d\rho_B}}- {\epsilon}={1\over3}{<:T_{ii}:>}
\end{equation}
and the Hugenholtz--Van Hove ($HV$) theorem \cite{huge58}:  
\begin{equation}
{\mu}={{d\epsilon}\over {d\rho_B}}=\epsilon_F
\end{equation}
are satisfied. Here $\epsilon = <:T_{00}:>$, $\rho_B = j^0$ is the 
baryon density and $\epsilon_F$ is the Fermi energy. 
This actually represents a good check of the consistency in the
truncation procedure used. We remind that the explicit inclusion
of meson field derivatives will make more delicate the proof
of the $HV$ theorem \cite{Ue89}.

In order to illustrate how the properties arising from the 
nucleonic intrinsic degrees of freedom can be described 
with our approach, we now present some results about the 
$EOS$ of asymmetric nuclear matter at equilibrium. 
The matrix ${\widehat F}_{\alpha\beta}(x,p)$ can be 
decomposed in the isospin space
as: ${\widehat F}_{\alpha\beta}(x,p)$ = 
$({\widehat F}_s)_{\alpha'\beta'}(x,p)   
+ \tau_3 ({\widehat F}_3)_{\alpha'\beta'}(x,p)$,
where $\tau_3$ is the third Pauli matrix, $(\alpha\beta)$ are spin-isospin
indices, while $(\alpha'\beta')$  denote spin indices. 
In symmetric nuclear matter the matrix ${\widehat F}_3(x,p)$ vanishes, 
whereas in the asymmetric case its components give
the various isovector densities. 
Due to the asymmetry of the system, the kinetic energy density increases 
because the Fermi momenta of protons and neutrons are different.      
The presence of the exchange terms (terms in square bracket in Eq.(11)) gives
rise also to a contribution from the potential part 
of the energy density 
$\epsilon$.

The symmetry energy can be defined as $E_{sym} = {\displaystyle 
{1\over{\rho_B}}{{\partial^2 \epsilon}\over {\partial I^2}}|_{I=0}}$, 
where $I$ is the asymmetry parameter, defined
as the ratio $\bigl((\rho_B)_n - (\rho_B)_p\bigr)/\rho_B$ 
($n$ stands for neutrons and $p$ for protons). In Fig.1 we have reported 
the calculated symmetry energy as a function of the baryon density
$\rho_B$. In the calculations the four 
parameters of the model ($f_S$, 
$f_V$, $A$ and $B$) have been fitted in order to reproduce equilibrium 
properties of {\it symmetric} nuclear matter:
saturation density $\rho_0=0.16 fm^{-3}$,
 binding energy E/A=-15.75 MeV, nucleon effective (or Dirac) mass at $\rho_0$,
${M_0}^*=0.7 M$, incompressibility modulus $K_0=250 MeV$.
In this way  we obtain the following value for the parameters:
$f_S=10.1$, $f_V=4.1$, $A= 0.06$ and $B= 0.01$.
In the scalar field contribution we clearly see a steady decrease
of the weigth of higher order terms. This seems to be a qualitative signal
of a {\it naturalness} in the expansion. However, a more accurate
discussion should be done only in a model that
includes non--linearly both scalar and vector fields
\cite{Fu96,Fu197}.

A quite interesting feature of our approach is that now 
the asymmetric nuclear matter is described relatively well with the
{\it same parameters} fixed by the fit. 
Figure 2 shows that our non--linear $HF$ (NLHF) evaluation (solid line) gives 
a symmetry energy value at normal
density, i.e. the coefficient of the Weiszaecker mass formula of 
$a_4 = 24~MeV$. This result appears too low with respect to the accepted
value (around $28-30MeV$), but we have to consider that  
only the contribution of isoscalar mesons are included. From this point
of view our calculation shows the importance of exchange
terms on the symmetry energy, even though the inclusion of the $\rho$ meson is 
necessary to achieve a good value of $a_4$.

To be more quantitative, considering that the kinetic contribution
to the $a_4$ parameter is of the order of $12MeV$ ($\simeq \epsilon_F/3$),
of the remaining part of  about $16MeV$ almost two thirds seem to come
from exchange terms.
This appears to be an interesting indication of a reduced contribution
of the $\rho$ meson strength in nuclear matter around normal density.

A rather repulsive density dependence is also deduced (Fig.2; full line). 
Non-linear Hartree (NLH) calculations including the $\rho$ meson (dashed line)
can reproduce the same $a_4$ value once the coupling constant $f_{\rho}$ is 
suitably fitted.
However in the $NLHF$ calculations we also obtain a non--linear density 
dependence of the symmetry energy 
that arises from the non--linear scalar self-interactions through
exchange terms.
This effect is not present in NLH calculations.
 

\section{Conclusions} 

We  have introduced a procedure to investigate the role 
of the Fock exchange terms in a relativistic model of fermions coupled 
to mesons, including self-interactions for the mesonic fields. 
We have shown that the evaluation of the Fock terms, within a 
suitable truncation
scheme, leads to a consistent description of the equilibrium and dynamical 
properties related to the fermionic intrinsic degrees of freedom. 
An application to the case of asymmetric nuclear matter at equilibrium 
gives satisfactory results concerning the density dependence of 
the symmetry energy. It should be noticed that also
other properties, such as the isospin and density dependence of effective 
masses, that can be deduced from the stationary solution of the kinetic
equation, are reasonably reproduced within our procedure, in comparison 
to QHD models including also scalar isovector mesons \cite{Kub97}.
A quite interesting general feature is appearing: a consistent 
inclusion of exchange contributions seems to clearly reduce
the weight of charged mesons in describing isospin properties 
of nuclear matter around and above normal barion density.

Finally we remark that the aim of this work is mainly to 
present a procedure
to evaluate Fock terms of a non--linear field coupling scheme.
For this reason we have omitted a full analysis 
of all possible features of a $EOS$ calculation; in particular our model
is subject to the possible criticism of all the approaches with non--linearities 
only in the
scalar field. However it is straightforward to extend the same procedure
to non--linear vector fields. 
Thus we stress that the method described here 
is very general and can be applied
to effective
field theories not only for nuclear matter but in general to 
systems with fermion-boson dynamical couplings.
Of course, the analysis can be extended also to the investigation of 
spin effects.


~\\[5ex]

{\bf Figure Caption}\\[5ex]
{\bf Fig.1}: 
Density behaviour of the coupling functions ${\tilde f}_S$ (bottom panel)
and ${\tilde f}_V$ (top panel). See the text for details.\\[2ex]

{\bf Fig.2}: 
Symmetry energy per nucleon vs. baryon density.
Full line: present $NLHF$ calculation. 
Dashed line: $NLH$ calculation including the
$\rho$ meson.


\begin{references}
\bibitem{newcomment}
In Re.[10] a HF approach to a non--linear model 
of QHD has been developed, however the Fock terms have been
evaluated linearizing the field equation for the scalar meson.
\end{references}
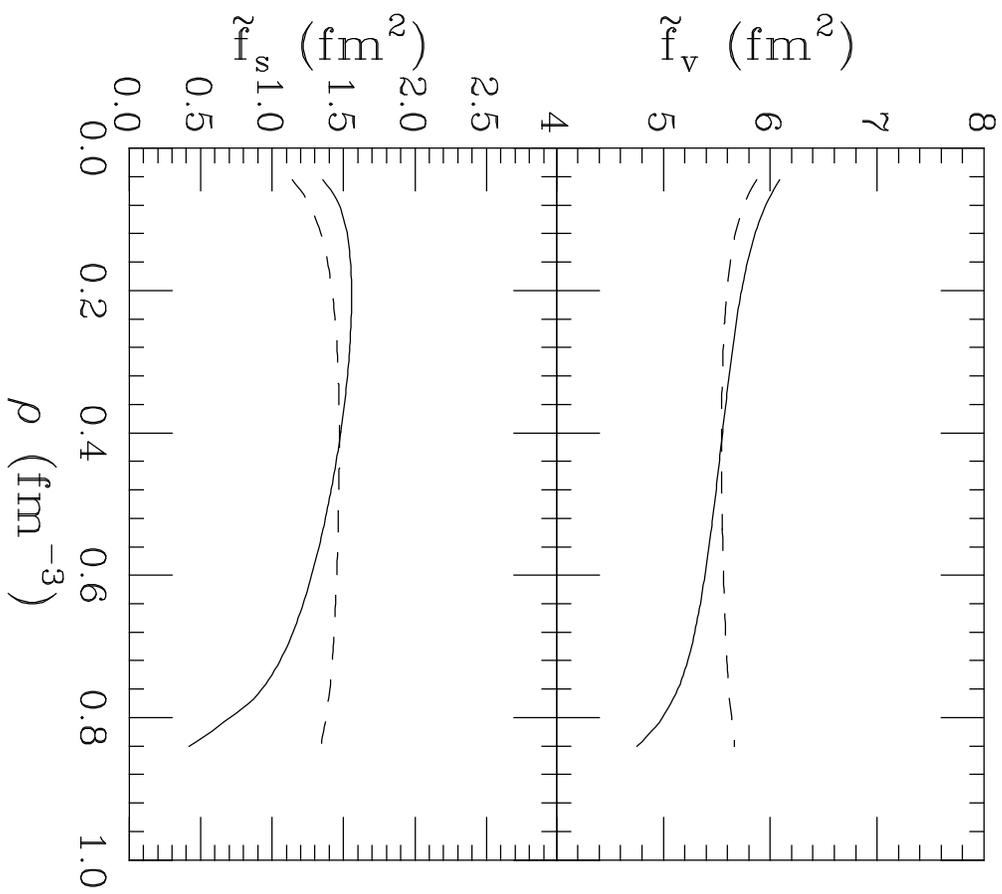
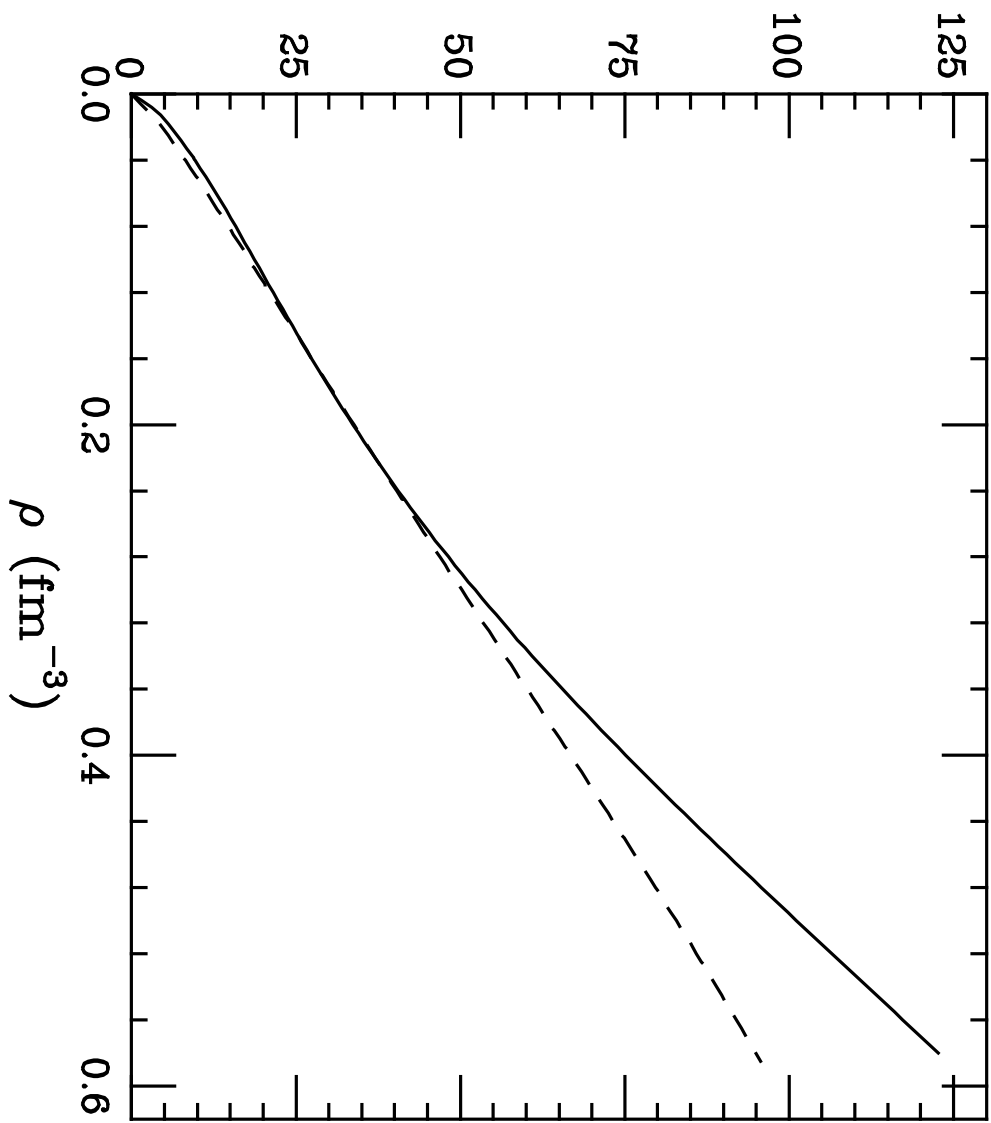
\end{document}